\documentclass{article}

\usepackage{amsfonts}
\usepackage{mathrsfs}
\begin{document}

\title{BRANEWORLDS AND QUANTUM STATES\\OF RELATIVISTIC SHELLS}

\author{S. ANSOLDI\\
{\small\it{}International Center for Relativistic Astrophysics (ICRA), Italy, and}\\
{\small\it{}Istituto Nazionale di Fisica Nucleare (INFN), Sezione di Trieste, Italy, and}\\
{\small\it{}Dipartimento di Matematica e Informatica, Universit\`{a} degli Studi di Udine,}\\
{\small\it{}via delle Scienze 206, I-33100 Udine (UD), Italy $\mathrm{[Mailing\ address]}$}\\
{\small\tt{}E-mail: ansoldi@trieste.infn.it}\\
{\small\tt{}Web-page: http://www-dft.ts.infn.it/$\sim$ansoldi}\\
\hbox to 9 cm {\hrulefill}\\
{\small{}To appear in the proceedings of the}\\
{\small{}Eleventh Marcel Grossmann Meeting on General Relativity,}\\
{\small{}July 23--29, 2006, Freie Universit\"{a}t Berlin, Berlin, Germany}\\
{\small{}({\tt{}http://www.icra.it/mg/mg11}).}\\
\hbox to 9 cm {\hrulefill}}

\maketitle

\begin{abstract}
We review some applications of relativistic shells that are relevant in the context of
quantum gravity/quantum cosmology. Using a recently developed approach, the stationary
states of this general relativistic system can be determined in the semiclassical
approximation. We suggest that this technique might be of phenomenological relevance
in the context of the brane-world scenario and we draw a picture of the general set-up
and of the possible developments.\\[3mm]
{{\it{}Keywords}: Brane World Scenario; Bohr--Sommerfeld States; Cosmology; General Relativistic Shells;
 Junction conditions; WKB Quantization.}
\end{abstract}

\bigskip\medskip\hrule\smallskip
\noindent{}Let us consider two (N+1)-dimensional domains of spacetime,
${}^{(N+1)}{\!}{\mathcal{M}} _{\pm}$, which
are parts of two solutions of Einstein equations;
let ${}^{(N)}{\!}\Sigma _{\pm}$ be isometric parts of their boundaries.
Then ${}^{(N)}{\!}\Sigma _{\pm}$ can be identified,
and ${}^{(N+1)}{\!}{\mathcal{M}} _{\pm}$
can be \emph{joined} across ${}^{(N)}{\!}\Sigma$. If
${}^{(N)}{\!}\Sigma$ is also equipped with some matter--energy content
and if it is a timelike hypersurface in
${}^{(N+1)}{\!}{\mathcal{M}}=${}%
${\,}^{(N+1)}{\!}{\mathcal{M}}_{-}{\,}\cup${}${\,}^{(N)}{\!}\Sigma\,\cup${}%
${\,}^{(N+1)}{\!}{\mathcal{M}} _{+}$,
then it describes the evolution of this matter/energy.
${}^{(N)}{\!}\Sigma$ is traditionally known as a \emph{general relativistic shell}
or a \emph{co-dimension one brane}.
General relativistic shells have been often used as a framework for
astrophysical and cosmological models (for an extensive bibliography
see Ref.~\cite{2002ClQGr..19....6321A}).
A good reason for this success is certainly the geometrically flavored description
provided by Israel junction conditions\cite{1966NuCiB..44..1.....I},
thanks to which the classical dynamics of the system is under control.
If we call ${}^{(N)}{\!}K ^{(\pm)} _{\mu \nu}$
($\mu, \nu = 1 ,${}$\dots{},${}$N+1$)
the extrinsic curvature of ${}^{(N)}{\!}\Sigma$ with respect
to its embeddings in ${}^{(N+1)}{\!}{\mathcal{M}} _{\pm}$ and ${}^{(N)}{\!}S _{\mu \nu}$ the stress--energy tensor
describing the matter--energy content of ${}^{(N)}{\!}\Sigma$, Israel junction conditions are,
in suitable units,
\begin{equation}
    {}^{(N)}{\!}K ^{(+)} _{\mu \nu} - {}^{(N)}{\!}K ^{(-)} _{\mu \nu} = 8 \pi M _{\mu \nu},
    \quad
    M _{\mu \nu} = {}^{(N)}{\!}S _{\mu \nu} - {}^{(N)}{\!}g _{\mu \nu} {}^{(N)}{\!}S / 2
    ,
\label{eq:juncon}
\end{equation}
where ${}^{(N)}{\!}g _{\mu \nu}$ is the metric on ${}^{(N)}{\!}\Sigma$ and ${}^{(N)}{\!}S$ is the trace of ${}^{(N)}{\!}S _{\mu \nu}$.
Soon after the earliest classical applications of shells, a number of works discussed
their semiclassical quantization
(see again Ref.~\cite{2002ClQGr..19....6321A} for additional
references). Most of them had the goal to investigate situations where
the emergence of singularities was breaking down the predictive power
of general relativity as in the cosmology of the early universe (with the initial
singularity problem) and in gravitational collapse (with its, also singular, final fate). In the first
case we would like to explicitly remember the paper of Farhi \emph{et al.}\cite{1990NuPhB..339.417...F},
which showed how useful the idea of shell tunnelling can be raising some
interesting (still open) issues\cite{PhReD2005..72103525J}.
About the second aspect, we remember the early works of Berezin\cite{1990PhLeB..241.194...B}
and Visser\cite{1991PhReD..43..402...V}
(additional bibliography can be found in Ref.~\cite{2002ClQGr..19....6321A}).
In what follows we will elaborate on the case in which\footnote{We will use the notation $\{{\mathscr{G}} _{\pm}\}$
to collectively indicate the dependence from the geometry--related parameters of the model (for
example, the Schwarzschild mass, the cosmological constant and so on) as
well as the notation $\{{\mathscr{E}}\}$ to denote the dependence from the parameters defining
the brane matter--energy content (for example, the surface tension and
so on). Later we will also use, with similar meaning, the shorthand $\{{\mathscr{G}}\}$ according
to the following definition:
$\{{\mathscr{G}}\} = \{{\mathscr{G}} _{+}\} \cup \{{\mathscr{G}} _{-}\}$.}
the metrics in ${}^{(N+1)}{\!}{\mathcal{M}} _{\pm}$ can be cast in the form
$
    {}^{(N+1)}{\!}d s ^{2} _{\pm}
    =
    - h _{\pm} (a _{\pm} ; \{{\mathscr{G}} _{\pm}\}) d t _{\pm} ^{2}
    + d a _{\pm} ^{2} / h _{\pm} (a _{\pm} ; \{{\mathscr{G}} _{\pm}\})
    + {}^{(N-1)}d \Omega ^{2} _{\pm} (\{{\mathscr{G}} _{\pm}\}) a ^{2} _{\pm}
$
in the coordinates $(t _{\pm},${}$a _{\pm},${}$(\dots) _{\pm})$, where ``$(\dots) _{\pm}$''
is a set of coordinates for the \emph{maximally symmetric spaces} of metric ${}^{(N-1)}d \Omega _{\pm} ^{2} (\{{\mathscr{G}} _{\pm}\})$.
In this setup the junction conditions (\ref{eq:juncon})
can be reduced to just one equation
\begin{equation}
    \epsilon _{+} \sqrt{\dot{A} ^{2} + h _{+} (A ; \{{\mathscr{G}} _{+}\})}
    -
    \epsilon _{-} \sqrt{\dot{A} ^{2} + h _{-} (A ; \{{\mathscr{G}} _{-}\})}
    =
    M (A ; \{{\mathscr{E}}\})
    ,
\label{eq:redjuncon}
\end{equation}
where $A (\tau)$ is the value of $a _{\pm}$ at the brane location as a function of the
proper time $\tau$ of on observer living on the brane, an overdot denotes a
derivative with respect to $\tau$ and
$\epsilon _{\pm}$ are the signs of the radicals.
$M (A ; \{{\mathscr{E}}\})$ encodes the properties
of the matter--energy source living on the shell.
Studies to develop the fully covariant Hamiltonian formulation started also
early\cite{1997PhReD..56..4706B} (see again Ref.~\cite{2002ClQGr..19....6321A}
for later ones)
and represent the proper framework to correctly interpret \emph{effective}
La\-gran\-gi\-an/Ha\-mil\-to\-ni\-an formulations on which, for simplicity, we will concentrate.
Indeed Eq.~(\ref{eq:redjuncon}) can be obtained from an effective, dimensionally
reduced Lagrangian/Hamiltonian as a first integral of the
Eu\-le\-r--La\-gra\-nge/Ha\-mil\-ton equations. From this Lagrangian/Hamiltonian,
following for instance Ref.~\cite{1997ClQuGr.14..2727..A}, the
effective momentum $P ( A , \dot{A} ; \{{\mathscr{G}}\} , \{{\mathscr{E}}\})$ conjugated
to $A$ can also be determined.
Moreover, from (\ref{eq:redjuncon}), it is possible to solve for $\dot{A}$ and substitute
this result into $P ( A , \dot{A} ; \{{\mathscr{G}}\} , \{{\mathscr{E}}\})$
to obtain $P ( A ; \{{\mathscr{G}}\} , \{{\mathscr{E}}\})$, i.e. an expression for the
\emph{momentum evaluated on a solution of the classical equations of motion}. If
the system admits bounded solutions, so that classically $A$ oscillates between
$A _{\mathrm{min}} (\{{\mathscr{G}}\} , \{{\mathscr{E}}\})$ and
$A _{\mathrm{max}} (\{{\mathscr{G}}\} , \{{\mathscr{E}}\})$, we can then evaluate
(sometimes analytically but, otherwise, at least numerically)
\emph{the value of the action on a classical solution}
\begin{equation}
    S (\{{\mathscr{G}}\} , \{{\mathscr{E}}\})
    =
    2
    \int _{A _{\mathrm{min}} (\{{\mathscr{G}}\} , \{{\mathscr{E}}\})} ^{A _{\mathrm{max}} (\{{\mathscr{G}}\} , \{{\mathscr{E}}\})}
        P (A ; \{{\mathscr{G}}\} , \{{\mathscr{E}}\}) dA
    .
\label{eq:claact}
\end{equation}
When the action $S (\{{\mathscr{G}}\} , \{{\mathscr{E}}\})$ is of the order of the quantum the
\emph{gravitational} system is in a \emph{quantum} regime and the Bohr--Sommerfeld quantization
condition
\begin{equation}
    S (\{{\mathscr{G}}\} , \{{\mathscr{E}}\}) \sim n \hbar , \quad n = 1 , 2 , \dots{},
\label{eq:bohsom}
\end{equation}
defines \emph{the semiclassical states of the system}. In this case
(\ref{eq:bohsom}) is a constraint:
\emph{not all combinations of values of the parameters are allowed}.
Let us now further specialize our discussion to $N = 4$
and discuss Robertson--Walker cosmologies in five--dimensional Schwarzschild
anti--de Sitter spacetime\cite{2000JHEP...09..014...D}
in the spirit of the Randall--Sundrum
scenario\cite{1999PhReL..83....3370S}.
Then $h _{+} (a ; \{{\mathscr{G}} _{+}\}) = h _{-} (a ; \{{\mathscr{G}} _{-}\}) = h (a ; \{k , l , m\} ) \equiv k + l ^{2} a ^{2} + 2 m / a ^{2}$
and $\epsilon _{+} = - \epsilon _{-} = +1$;
we also choose the coordinates in the maximally symmetric space as
$(\dots) _{\pm} = (\chi _{\pm}, \theta _{\pm}, \phi _{\pm})$. Then
$
    {}^{(3)}d \Omega ^{2} _{\pm} (\{{\mathscr{G}} _{\pm}\})
    =
    {}^{(3)}d \Omega ^{2} _{\pm} (k)
    =
    d \chi ^{2} _{\pm}
    + f _{k} ^{2} (\chi _{\pm})
        ( d \theta _{\pm} ^{2} + \sin ^{2} \theta _{\pm} d \phi _{\pm} ^{2})
$,
where
$
    f _{k} (y)
    =
    (\exp (k ^{1/2} y) - \exp (- k ^{1/2} y))/(2 k ^{1/2} y)
$
and
$k = -1,0,+1$ determines if the maximally symmetric space is a $3$-sphere, ${\mathbb{S}} ^{3}$,
a $3$-Euclidean space, ${\mathbb{E}} ^{3}$,
or $3$-Hyperbolic space, ${\mathbb{H}} ^{3}$, respectively.
We now observe that in the model we are discussing, $M (R ; \{{\mathscr{E}}\})$
contains most of the relevant physical information, since it describes the
matter--energy content of the brane, i.e. of our universe. It is, then,
interesting to choose the set of parameters $\{{\mathscr{E}}\}$ to describe the matter
component of the universe $\rho _{\mathrm{m}}$, the radiation component $\rho _{\mathrm{r}}$,
the cosmological constant $\rho _{\Lambda}$ the dark energy $\rho _{?}$ and so on;
thus $\{{\mathscr{E}}\}=\{ \rho _{\mathrm{m}} , \rho _{\mathrm{r}} , \rho _{\Lambda} , \rho _{?}\}$,
whereas $\{{\mathscr{G}}\} = \{ k , l , m \}$ is fixed by the bulk spacetime
structure. We then see that, already in the very simple and natural semiclassical approach discussed
above (of which the toy model in Ref.~\cite{205AIPcp.751...159A} is a preliminary test), the
quantization condition
$
    S ( \{ k , l , m \} , \{ \rho _{\mathrm{m}} , \rho _{\mathrm{r}} , \rho _{\Lambda} , \rho _{?} \} ) \sim n \hbar
$, $n = 1 , 2 , \dots{}$,
provides a constraint among the cosmological parameters.
Phenomenological implications and further refinements of this approach
are currently under investigation and will be reported elsewhere\cite{2005InPre......AnGuIsA}.

\section*{Acknowledgements.}

I would like to thank Mr. Bernardino Cresseri, Prof. Gianrossano Giannini
and Mr. Enrico Ramot for some administrative and practical arrangements which
made possible my participation to the MG11 meeting. I would also like to
gratefully acknowledge financial support from ICRA (International
Center for Relativistic Astrophysics) and INFN (Istituto Nazionale
di Fisica Nucleare, Sezione di Trieste).

\end{document}